\documentclass{aa}  
\usepackage{graphicx}
\usepackage{txfonts}

\def\bc{{\sc {BRITE-Constellation}}}
\def\b{{\sc {BRITE}}}

\def\deg{$^{\circ}$}

\def\ac{$\alpha$\,Cir}

\def\cd{\rm d$^{-1}$}

\def\w{{\sc \small{WIRE}}}

\def\arcsec{\hbox{$^{\prime\prime}$}}

\newcommand{\muHz}{\mbox{$\mu$Hz}}

\begin{document} 

\title{The roAp star \ac\ as seen by BRITE-Constellation\thanks{Based on data collected by the \bc\ satellite mission, built, launched and operated thanks to support from the Austrian Aeronautics and Space Agency, the University of Vienna, the Canadian Space Agency (CSA),  the Foundation for Polish Science \& Technology (FNiTP MNiSW), and National Centre for Science (NCN). \break $\dagger$ Member of the BRITE Executive Science Team (BEST). \hfill \break $\diamond$ Member of the Photometry Tiger Team (PHOTT).}} 

\author{W.\,W. Weiss \inst{1 \dagger \diamond }
\and H.-E. Fr\"{o}hlich\inst{2}
\and A. Pigulski \inst{3 \dagger \diamond}      
\and A. Popowicz \inst{4 \dagger \diamond}      
\and D. Huber\inst{8}           
\and R. Kuschnig\inst{1 \dagger \diamond }
\and A.\,F.\,J.\,Moffat\inst{5 \dagger \diamond}
\and J. M. Matthews\inst{6 \dagger}             
\and H.  Saio\inst{10}                  
\and A. Schwarzenberg-Czerny\inst{11 \dagger \diamond}     
\and C. C. Grant\inst{7}                        
\and O. Koudelka\inst{9\dagger}         
\and T. L\"uftinger\inst{1}             
\and S.\,M. Rucinski\inst{12 \dagger \diamond}          
\and G. A. Wade\inst{15 \dagger \diamond}       
\and J. Alves\inst{1 \dagger}
\and M. Guedel\inst{1 \dagger} 
\and G. Handler\inst{11 \dagger \diamond} 
\and St. Mochnacki\inst{12 \dagger \diamond}  
\and P. Orleanski \inst{13 \dagger}   
\and B. Pablo\inst{5 \diamond}          
\and A. Pamyatnykh\inst{11 \dagger}    
\and T. Ramiaramanantsoa\inst{5 \diamond}       
\and J. Rowe\inst{14}   
\and G. Whittaker\inst{7 \diamond}              
\and T. Zawistowski\inst{13 \diamond}   
\and E. Zoc{\l}o\'nska\inst{11 \diamond}  
\and K. Zwintz\inst{16 \dagger}
        }   
                        
\institute{University of Vienna, Institute for Astrophysics, Vienna, Austria; \email{werner.weiss@univie.ac.at}        
\and Kleine Strasse 9, D-14482 Potsdam;  Leibniz-Institut f\"ur Astrophysik (AIP), Potsdam, Germany;  \email{hefroehlich@aip.de}                                                                                                                            
\and Astronomical Institute, University of Wroc{\l}aw, Poland;  \email{pigulski@astro.uni.wroc.pl}                      
\and Institute of Automatic Control, Silesian University of Technology, Gliwice, Poland; 
\and Dept. de physique, Universit$\acute e$ de Montr$\acute e$al; 
\and Dept. of Physics and Astronomy, University of British Columbia, Canada; 
\and Space Flight Laboratory, University of Toronto; 
\and Sydney Institute for Astronomy, University of Sydney, NSW 2006, Australia; 
        SETI Institute, Mountain View, CA 94043, USA;                                                                           
        Stellar Astrophysics Centre, Aarhus University, Aarhus C, Denmark                                                         
\and Graz University of Technology, Austria; 
\and Astronomical Institute, Graduate School of Science, Tohoku University, Sendai, Japan; 
\and Nicolaus Copernicus Astronomical Center, Warsaw, Poland; 
\and Dept. of Astronomy and Astrophysics, University of Toronto, Canada; 
\and Space Research Center, Warsaw, Poland; 
\and NASA Ames Research Center; 
\and Dept. of Physics, Royal Military College of Canada, Ontario, Canada; 
\and Institute for Astro- and Particle Physics, University of Innsbruck; 
        }
\titlerunning{roAp star \ac}
\authorrunning{Weiss, Fr\"{o}hlich, Pigulski et al.}
\offprints{werner.weiss@univie.ac.at}

\date{Received   / Accepted }

\abstract{
We report on an analysis of high-precision, multi-colour photometric observations of the rapidly-oscillating Ap (roAp) star \ac . These observations were  obtained with the \bc, which is a coordinated mission of five nanosatellites that collects continuous millimagnitude-precision photometry of dozens of bright stars for up to 180 days at a time in two colours ($\approx$ Johnson B and R). \b\ stands for BRight Target Explorer. The object \ac\  is the brightest roAp star and an ideal target for such investigations, facilitating the determination of oscillation frequencies with high resolution. This star is bright enough for complementary interferometry and time-resolved spectroscopy. Four \b\
satellites observed \ac\ for 146\,d or 33 rotational cycles.
Phasing the photometry according to the 4.4790\,d rotational period reveals qualitatively different light variations in the two photometric bands. The phased red-band photometry is in good agreement with previously-published WIRE data, showing a light curve symmetric about phase 0.5 with a strong contribution from the first harmonic. The phased blue-lband data, in contrast, show an essentially sinusoidal variation. We model both light curves with Bayesian Photometric Imaging, which suggests the presence of two large-scale, photometrically bright (relative to the surrounding photosphere) spots. We also examine the high-frequency pulsation spectrum as encoded in the \b\ photometry. Our analysis establishes the stability of the main pulsation frequency over the last $\approx 20$\,years, confirms the presence of frequency $f_7$, which was not detected (or the mode not excited) prior to 2006, and excludes quadrupolar modes for the main pulsation frequency.
}
   \keywords{Asteroseismology -- Stars: chemical peculiar, oscillation, rotation, spots, fundamental parameters -- Stars individual: \ac\ (HR 5463, HD 128898)}
   \maketitle
   

\section{Introduction}  
 \ac\ (HD\,128898, HR\,5463) is a binary system whose primary component is an A7VpSrCrEu star (V\,=\,3.19) with a K5V companion (V\,=\,8.47) separated by 16\arcsec \ \citep{way}. (The companion is faint enough and distant enough from the primary to be ignored in the following study.)  The primary component of \ac\ was discovered as the fifth known rapidly oscillating Ap (roAp) star in 1981 by \citet{kurtz1}, followed by a frequency analysis of 47 hours of high-speed, ground-based photometry \citep{kurtz2}. These authors pointed to a variation in the amplitude of the principal pulsation, which is suspected to be due to rotational modulation.  Pulsation of \ac\ was confirmed by \citet{schnei1} and soon after, \citet{weiss1} presented  the first time-resolved simultaneous multi-colour photometry of \ac\ in the Walraven system. The goal was pulsation mode identification (attempted also by \citet{kurtz3}) based on frequency multiplets and amplitude changes. 

The first simultaneous time-resolved photometry and spectroscopy of \ac\ were obtained at ESO by \citet{schnei2}. This yielded only marginal evidence for radial velocity variations for two Ca\,I lines and the authors remarked on the very uncertain value on the rotation period for this star (ranging in the literature from one to 12 days). Five years later, \citet{kurtz4} provided a robust value of the rotation period (P $\simeq$ 4.46\,d). Another 15 years elapsed before the first light curve sampling the full rotation cycle of \ac\ was published based on \w\ space photometry  \citep{wire1}.

The quantity and quality of data, and the sophistication of models, has improved dramatically over the subsequent years (e.g. \citet{kurtz4}; \citet{audard}; \citet{balona}; \citet{koch1}; and many references therein). The most complete and recent summary of our knowledge about the pulsation of \ac\ is presented by \citet{wire2,wire1} 
and of the complex atmospheric structure and spectral signatures of pulsation by \citet{ryab} and \citet{koch2}.

\ac\ was included in proposals submitted in 2009 and 2010 by several authors of this paper, responding to an Announcement of Opportunity for targets to be observed by \bc. Proposal \#05 (WW) focused on roAp stars, \#19 (DH) on 50 bright stars with the most accurately known {\sl Hipparcos} parallaxes, and \#25 (TL) on the rotation of chemically peculiar stars (in particular, possible changes in rotation period). 


\section{BRITE-Constellation photometry}                        \label{s:phot}      

The \bc\ is a coordinated mission of five nanosatellites located in low Earth orbits, each hosting an optical telescope of  3\,cm aperture feeding an uncooled CCD, and observing selected targets in a 24$^{\circ}$ field of view \citep{brite1}. Each nanosat is equipped with a single filter; three have a red filter (${\sim}$\,620\,nm) and two have a blue filter (central wavelength (${\sim}$\,420\,nm).  The satellites have overlapping coverage of the target fields to provide two-colour, time-resolved photometry.

The detector is a KAI-11002M CCD with 11 million $9\times 9\,\mu$m pixels, a 14-bit A/D converter, gain of about 3.3\,e$^-$/ADU, and a dark current of 1e$^-$/s at $+20\degr $C. The saturation limit of the pixels at this temperature is $\sim 13\,000$\,ADU, and the response is linear up to about 8\,000\,ADU. The pixel size is 27.3\arcsec , projected at the sky.

The five nanosats are designated \b -Austria (BAb), Uni\b\ (UBr),  \b -Lem (BLb), \b -Hewelius (BHr), and \b -Toronto (BTr), where the last letter indicates the filter on board the satellite ("b" = blue, "r" = red). A sixth nanosat, \b -Montr\'eal (BMb), did not detach from its launch vehicle. The nanosats Bb and UBr were launched first, followed by BLb, BTr, and BHr. 

\subsection{\ac\ photometry}

\ac\ was one of the targets in the Centaurus field, the second long-term pointing of \bc\ after the Orion field.  The early data were obtained during science commissioning of the mission, which meant interruptions in science data collection for engineering tests.  Even so, significant gaps in the time series were infrequent, caused mainly by accidental spacecraft mode changes from fine-to-coarse pointing (or tumbling) with recovery times shorter than 48 hours.  New nanosats joined the Constellation during the \ac\ run. As a consequence of this dynamic, instructional operational environment, the \ac\ data set is naturally somewhat complicated.

The data cover 25 March -- 8 August 2014: 146\,days, corresponding to nearly 33 stellar rotational cycles (see the next Sec.\,\ref{s:rot} for details). 

\ac\ was observed in the blue by BAb for 131 days and BLb for 26 days; in the red by UBr for 145 days, and BTr for six days.  The light curves consist of nearly 44\,000 blue exposures and 69\,000 red exposures, each of duration 1\,s in both colours.
The sampling time was 21.3\,s, which is set by the amount of processing time needed between consecutive exposures and the maximum amount of data that could be stored on board owing to limited download capacity at the time. 

\ac\ was observed during individual \b\ orbits for 5 to 30\,min, depending on target visibility, the Earth Exclusion Angle (angular distance between the illuminated limb of the Earth and the star-tracker field of view), the required warm-up period for the star trackers after reacquisition of the target, and passages of the satellites through the South Atlantic Anomaly (when higher fluxes of cosmic rays prevent useful signal from the star trackers).  There were also brief interruptions of science data collection when data were being transmitted to ground stations.

The daily data transfer rate to ground stations in Europe was less than planned   because of heavy interference with unidentified transmitters on the ground, which is particularly serious during the Centaurus field observations. For example, data transfer from BAb to the station in Graz, Austria, was limited to a maximum of 15\,min per orbit.

\subsection{Reduction: \b\ standard pipeline}          \label{ss:pipe}

During the commissioning of {\bc}, the particular challenges of on-orbit photometry were assessed. To develop an efficient pipeline for fast and optimum extraction of the photometry, a Photometry Tiger Team (PHOTT) was established, consisting of three groups working independently and pursuing different processing and reduction strategies. 

Issues investigated included: CCD radiation defects (warm columns, hot and cold pixels), CCD temperature variations, charge-transfer inefficiencies, pixel-to-pixel sensitivity variations and intra-pixel inhomogenities, pointing errors, complex point spread functions (PSF) varying over the 24$^{\circ}$ field of view and during long observing runs, identification of defective images (e.g. incomplete PSF in a sub-raster). Competing reduction methods employed PSF fitting and aperture photometry
as well as various threshold criteria for elimination of bad data points (outliers), various fit-quality parameters, median vs. mean, etc. The methods will be described in detail by \citet{phott1}.

\begin{figure}[h]               
\includegraphics[width=0.49\textwidth]{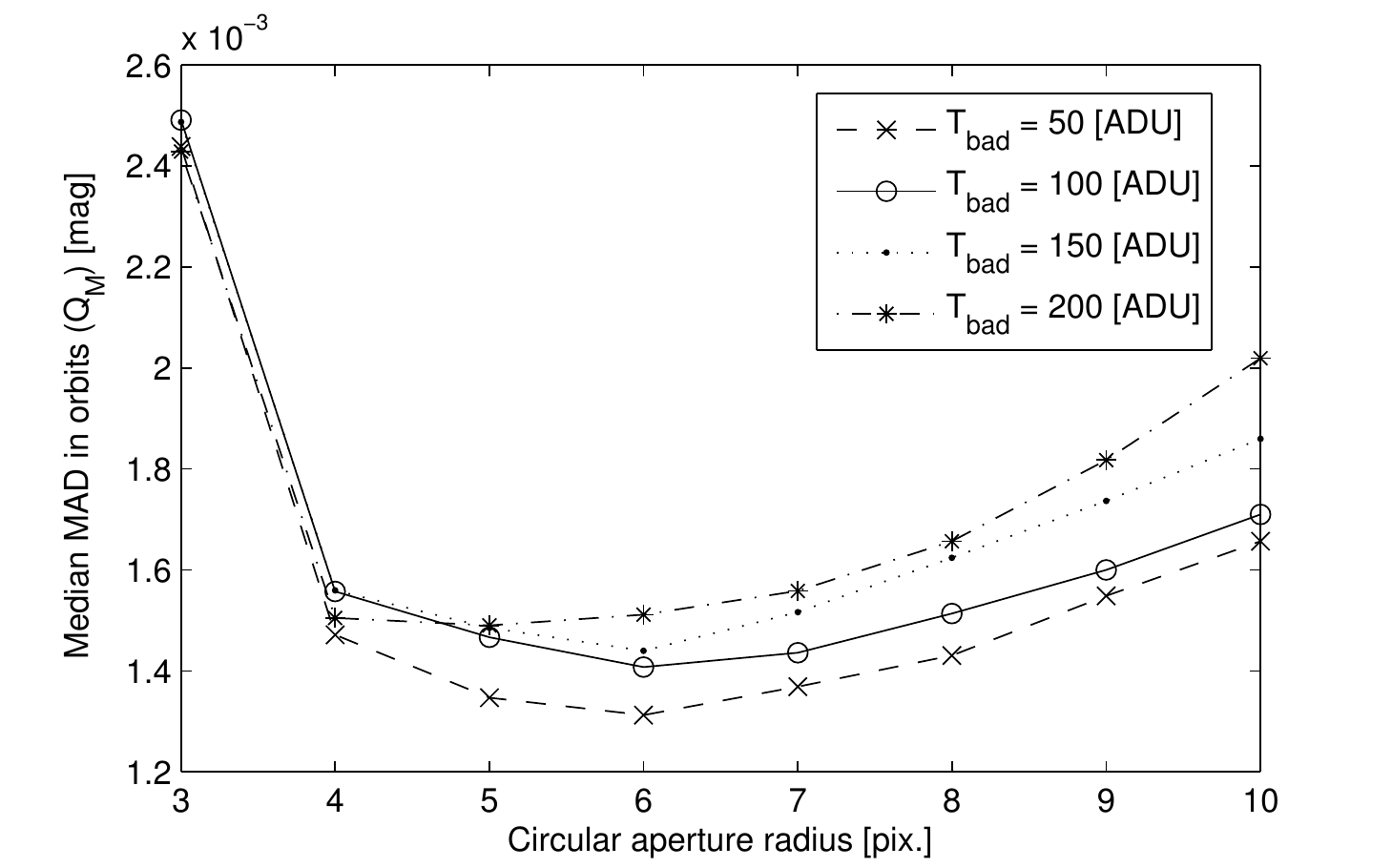}
\caption{Optimising pipeline parameters for UBr data: bad pixel elimination ($T_{\rm bad}$) and aperture radius.}
\label{optim} 
\end{figure}

This investigation of \ac\ is based on the reduction pipeline  developed by PHOTT member Adam Popowicz via aperture photometry. Figure\,\ref{optim} illustrates the impact of two optimisation parameters on the light-curve quality. The parameters are the threshold employed for bad-pixel identification in the mean dark frame ($T_{\rm bad}$) and the aperture radius. To assess the light-curve quality, we computed the median absolute deviations (MAD) of results obtained for individual satellite orbits, as a robust measure of scatter in the presence of outliers. The final quality measure ($Q_{\rm M}$) is the median of MADs defined as follows:

\begin{equation}
      Q_{\rm M} = \textrm{median}\left\{\frac{\textrm{MAD}_i}{\sqrt{N_i}}\right\} ,
\end{equation}
where $\textrm{MAD}_i$  and $N_i$ denote the median absolute deviation and the number of measurements in the $i$-th orbit with a total number of orbits $N_O$.

\begin{figure}[h]       
\includegraphics[width=0.49\textwidth]{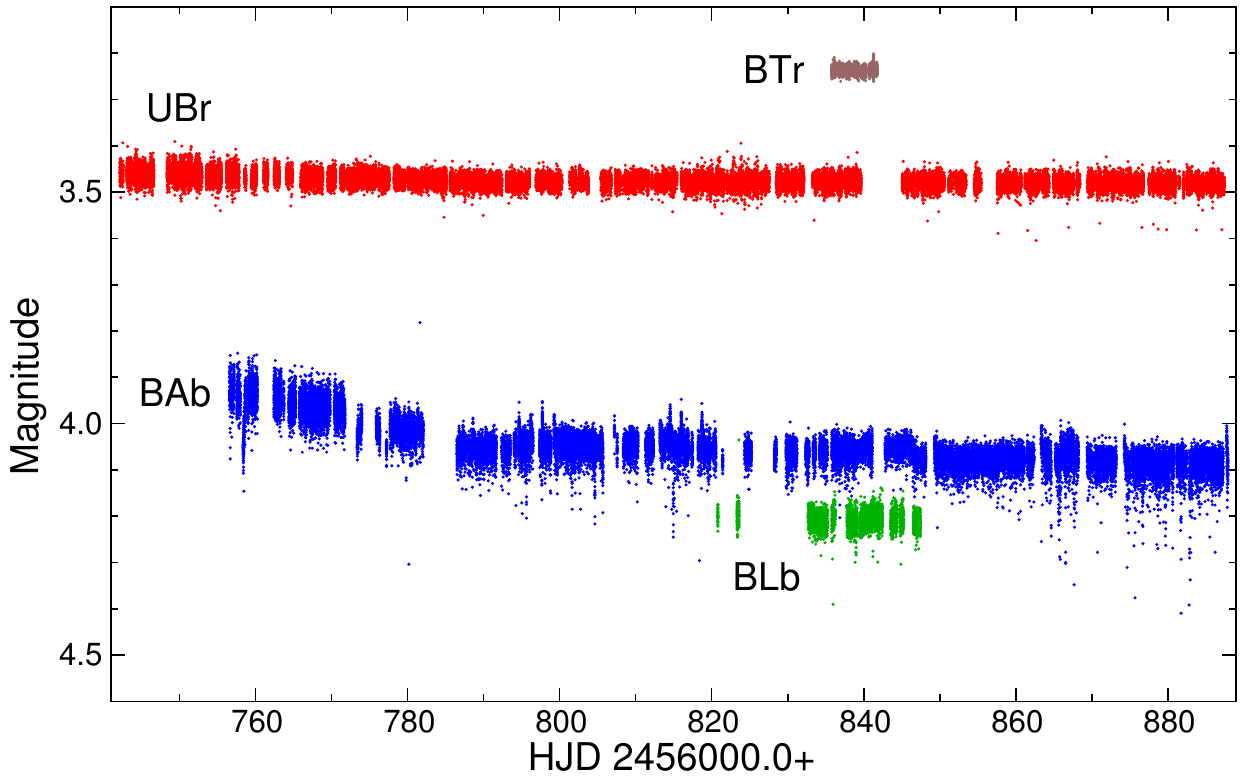}
\caption{Pipeline light curves of \ac\ for the four \b s. Note offsets for BTr and BLb and the larger scatter for the blue data.}
\label{LCo}  
\end{figure}

Optimisation is the most time-consuming stage of this pipeline, so only four values for $T_{\rm bad}$ (50, 100, 150, and 200 ADU) and eight values for the aperture (ranging from 3 to 10 pixels) were adopted. This seems sufficient to optimise parameters for the \ac\ data (see minima in optimisation curves presented in Fig.\,\ref{optim}).  The final pipeline light curves from the four \b s that observed \ac\ are shown in Fig.\,\ref{LCo}.

\subsection{Additional reductions}      \label{ss:tune}    

Figure\,\ref{LCo} reveals that the pipeline reduction does not produce a truly clean light curve in this case, and that some
outliers and instrumental effects have survived the pipeline. 
The pipeline data were fine-tuned in three steps:  (i) removal of outliers, (ii) decorrelations, and (iii) removal of the \b\ satellite orbits that returned the poorest quality of photometry. (More details on fine-tuning \b\ pipeline-reduced data can be found in the BRITE Data Analysis Cookbook at http://brite.craq-astro.ca/doku.php?id=cookbook.)

\begin{figure}[h]       
\includegraphics[width=0.49\textwidth]{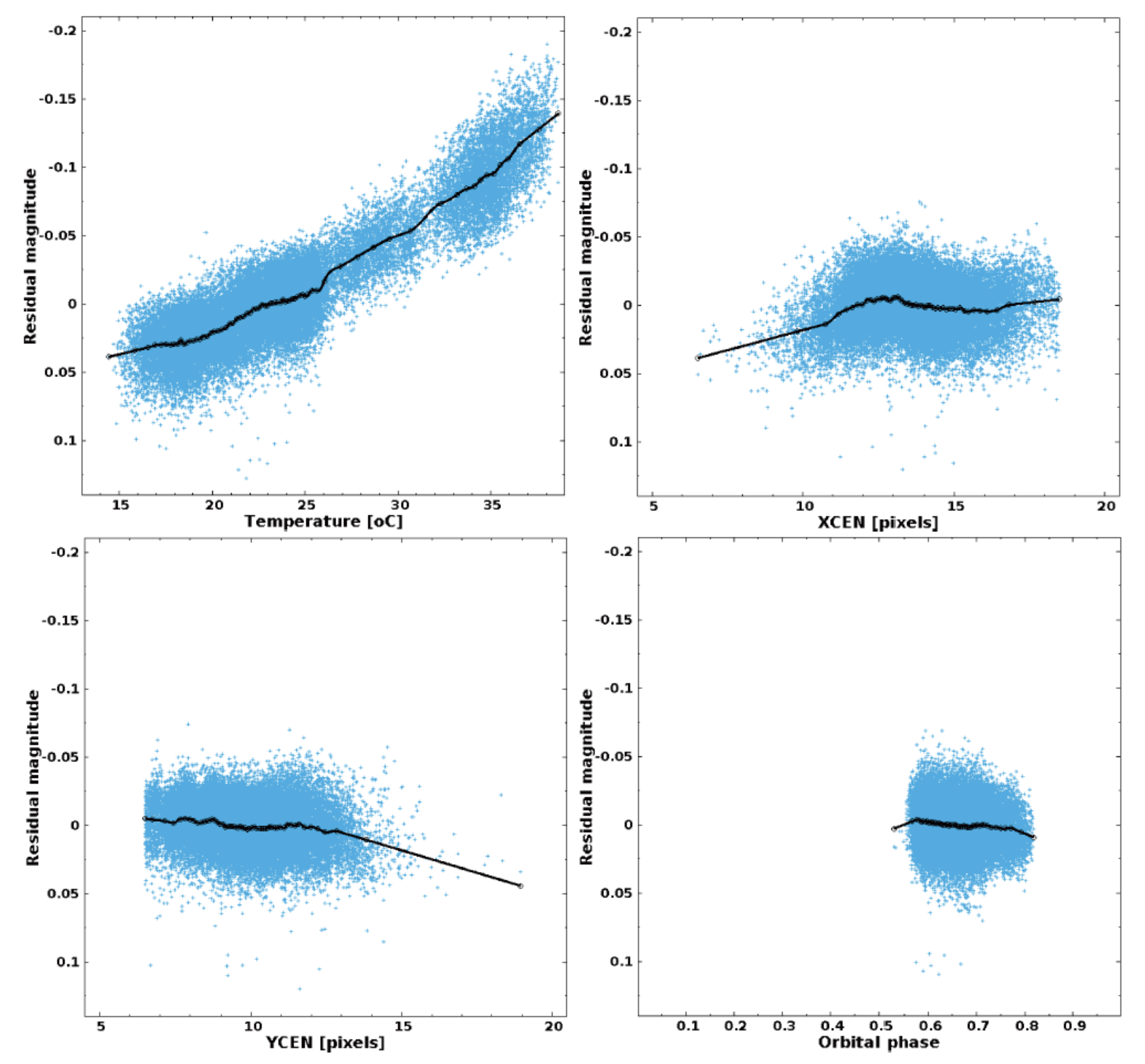}
\caption{Correlations between residual magnitudes and four instrumental parameters of BAb. The continuous line follows the Akima interpolation of bin averages.}
\label{Decorr}   
\end{figure}

First, outliers were eliminated based on a Generalized Extreme Studentized Deviate (GESD) test \citep{rosner} applied to data obtained during a single orbit of each satellite. The number of identified outliers depends on a single parameter $\alpha$, increasing with larger $\alpha$.


While removing more outliers results in smaller scatter, it also results in fewer data points. Consequently, the signal detection threshold in a periodogram, which depends both on the number of data points and their scatter, can have a minimum as a function of $\alpha$. We adopted optimum values of $\alpha\,=\,0.8$ for BAb and UBr data, 0.5 for BLb, and 0.3 for BTr data, accepting a small increase in noise level.  The strongest intrinsic stellar variations (the rotation frequency $f_{\rm rot},\  2f_{\rm rot}$ and the main pulsation frequency near 6.5\,min, $f_1$) were filtered from the data prior to the outlier removal.

\begin{figure}[b]       
\includegraphics[width=0.49\textwidth]{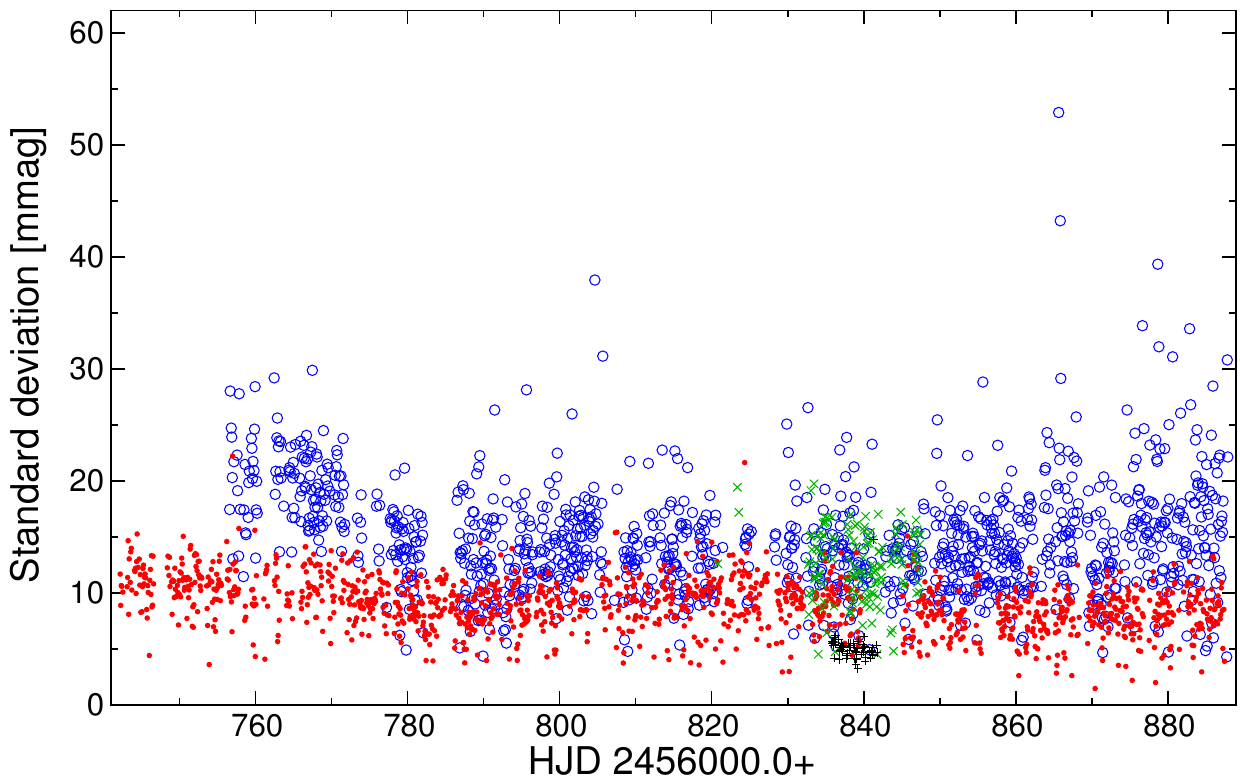}
\caption{Standard deviations for individual orbits: BAb (open circles), UBr (dots), BLb (x), BTr (+)}
\label{stdev}   
\end{figure}

In the next step, we decorrelated the data with temperature, followed by a decorrelation with the position of the stellar centres in both coordinates and orbital phase, as is illustrated in Fig.\,\ref{Decorr} for BAb data. After each decorrelation, the outlier removal was repeated with the previously used $\alpha$ parameter. 

\begin{figure}  
\includegraphics[width=0.49\textwidth]{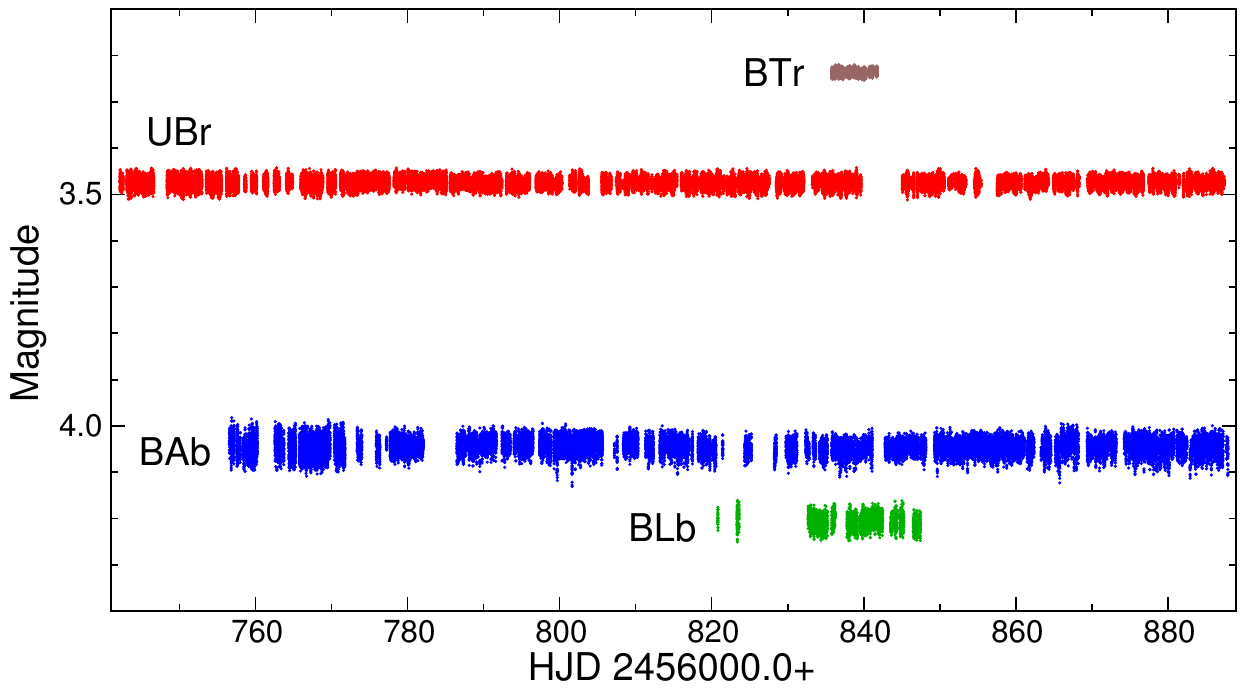}
\caption{\ac\ light curves after removal of the worst orbits and outliers and decorrelation. BTr and BLb data are offset for better visibility.}
\label{LCf}   
\end{figure}

The next step towards the final light curve was rejection of the worst orbits, based on a single threshold for a given \b . Although it may not be obvious in Fig.\,\ref{LCo}, there are few orbits that returned photometry of poor quality. This is better seen in Fig.\,\ref{stdev}, which shows the standard deviations in mmag for individual orbits. Based on this, the chosen thresholds are: 26\,mmag for BAb, 25\,mmag for BLb, 17\,mmag for UBr, and 8\,mmag for BTr.

Comparing Fig.\,\ref{LCf} with Fig.\,\ref{LCo} illustrates the improvements achieved by fine-tuning the original pipeline data. The systematic decrease of BAb magnitudes (for example, early in the observing run and seen in Fig.\,\ref{LCo}) disappeared because it was caused by CCD temperature variations (Fig.\,\ref{Decorr}, upper left). After fine-tuning, a total of 61\,988 red and 34\,592 blue data remain.  

\section{Rotation of \ac }                              \label{s:rot}    

The first indirect evidence of the rotation period of \ac\ was reported by \citet{kurtz4} who deduced a rotation period of about 4.46 days from the amplitude modulation of the star's main pulsation mode.  This initial estimate was refined to 4.4790 days (Kurtz, op.cit) through photometry collected over the following 12 years from various observing campaigns. \ac\ was observed by the \w\ satellite star tracker for a total of 84 days during September 2000 -- July 2006, supported by ground-based Johnson B photometry from SAAO. These data led to the first rotation light curve of \ac\ \citep{wire1}.

\begin{figure}[h]       
\includegraphics[width=0.5\textwidth]{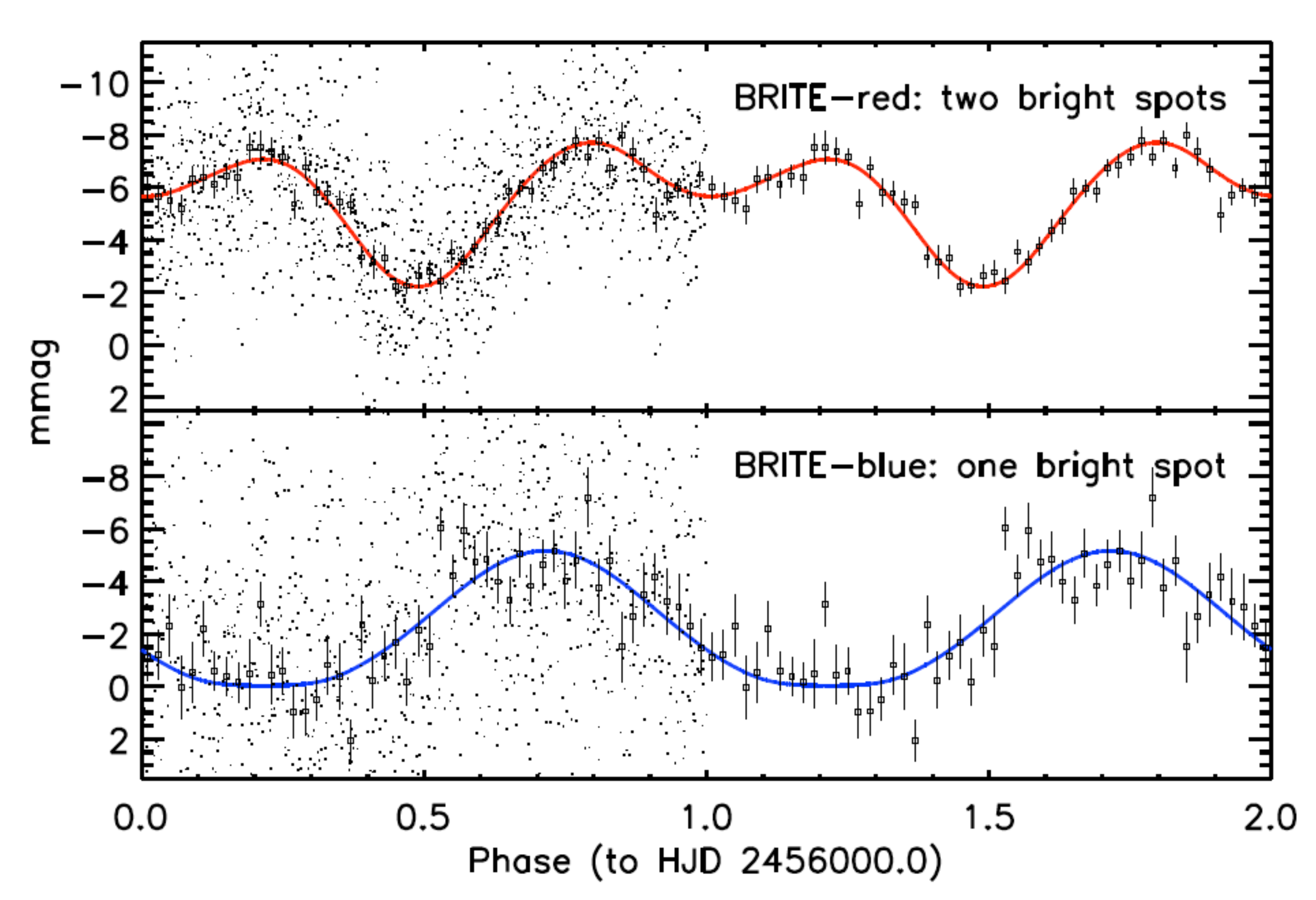}
\caption{Rotation of \ac\ observed by the red-sensitive (top) and blue-sensitive  (bottom) \b s, respectively, with P$_{rot}$\,=\,4.4790$^d$. Phase-bin averages with 1-$\sigma$ errors are given together with the theoretical light curve obtained for the respective spot model (see Sec.\,\ref{s:bpi}). Phases 0 to 1 also show the binned original \b\ data with an overall r.m.s. noise of 9\,mmag for the red and of 15\,mmag for the blue light curve.}
\label{comcol}   
\end{figure}

\bc\ takes the analysis of \ac\ to a new stage compared to \w\ by providing space-based photometry in two colours. The photometry is plotted in phase with the rotation period in Fig.\,\ref{comcol}. The noise level for the blue data is larger, mostly due to problems with the BAb star tracker and as a result of the spectral type, which produces a smaller count rate for the blue \b s compared to the red BRITEs. The rotation light curves  are obviously different in the two
colours. While one maximum at approximately phase $\phi = 0.8$ appears in both filters, the second maximum in red at about $\phi = 0.2$ is not present in blue (see Fig.\,\ref{comcol}). This difference is significant, likely indicating very peculiar spot properties and posing a challenge to interpret physically.

\begin{figure}[h]       
\includegraphics[width=0.49\textwidth]{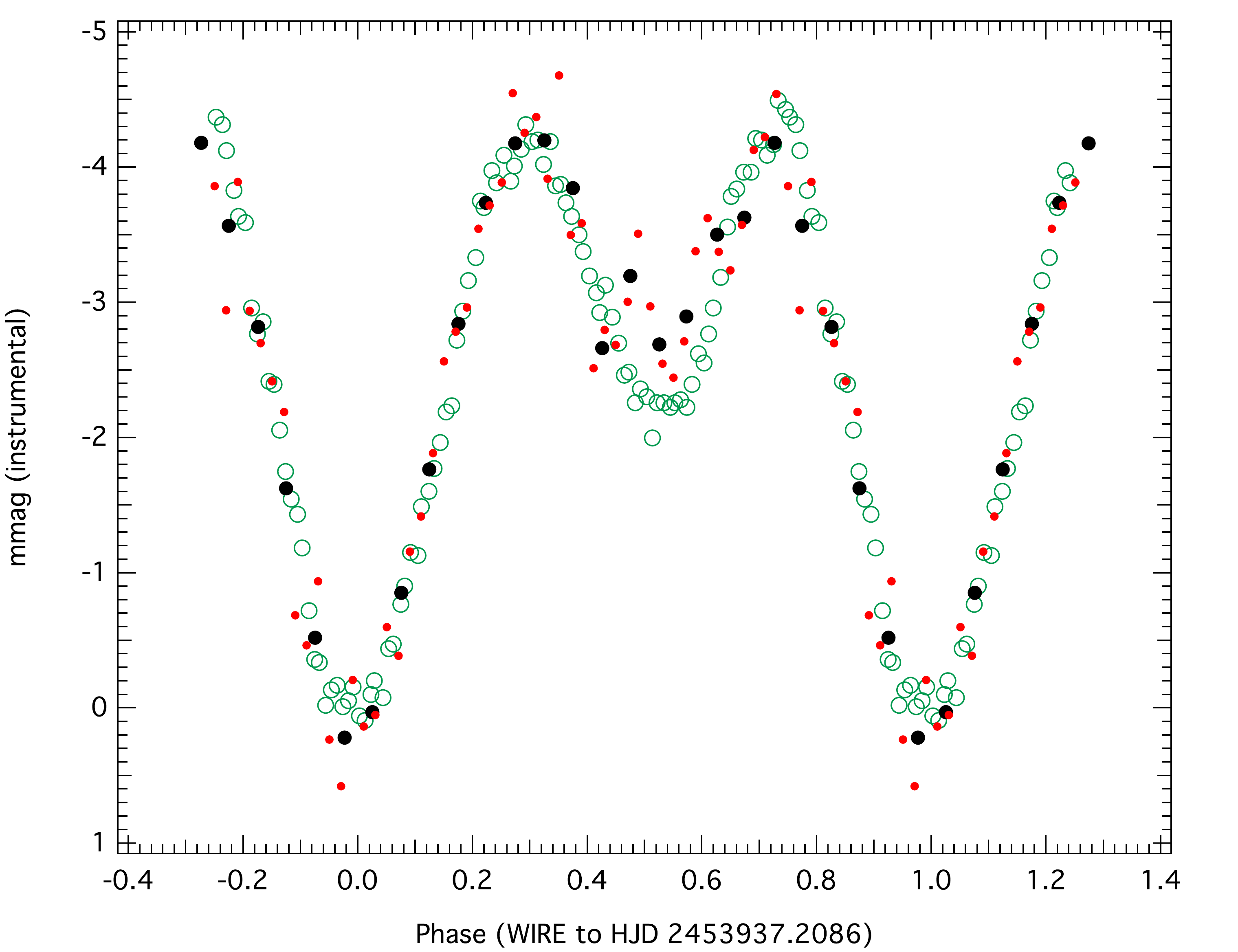}
\caption{Rotation of \ac. Green circles: \w\ - 2006 July data (Fig.\,2 of \citet{wire1}). Dots: UBr and BTr data (red filter) with a phase binning of 0.01 (small  red dots) and of 0.05 (large black dots). The \b\ data were scaled to fit the amplitude range of \w\ and shifted in phase to be compatible with the different reference time in HJD for \w\ and \b .}
\label{comprot}   
\end{figure}

For comparison, we digitised Fig.\,2 of \citet{wire1} and, in our
Fig.\,\ref{comprot}, plotted  their 100 \w\ data points from July 2006  along with our \b\ red data (similar in bandpass to the \w\ data). We conclude that the rotational modulation pattern has been stable between 2006 and 2013/14. When comparing the scatter, one should keep in mind that (1) \bc\ data are based on 1\,s integrations sampled every 21\,s, while \w\ data were obtained twice per second, co-added in 15\,s bins; and (2) the collecting area of the \w\ star-tracker telescope is three times larger than the \b\ telescope.

\section{Bayesian photometric imaging (BPI)}    \label{s:bpi}         

Ap stars, such as \ac,\ are well known to exhibit surface abundance nonuniformities (abundance spots) produced as a consequence of chemical diffusion in the presence of their strong magnetic fields. As described by, for example \citet[][]{krt0}, photometric variability of these stars is believed to result from flux redistribution  principally because of variable line blanketing modulated by stellar rotation. \ac\ exhibits obvious photometric variability on the rotational timescale that we interpret to be indicative of such spots.

\subsection{Likelihood estimate}

The Bayesian photometric imaging (BPI) technique to construct a spot model based on a light curve well sampled in time is described in detail by \citet{luefta}. Below we only comment on the construction of the likelihood function, vital for any parameter estimation. 

To save computation time, the fine-tuned data were combined into bins with a maximum width of 36\,min. This bin width corresponds to roughly one-third of the orbital period of one \b\ satellite. The time difference between adjacent points within a bin does not exceed 15\,min.
With these chosen bin constraints, no bin contains fewer than ten data points. Each of the 1368 red and 1200 blue bins was assigned a weight according to the number of contributing data points within a bin and used for the BPI procedure described below. 

We assume the residuals, deviations between observation and model, are Gaussian-distributed, i.e. measurement errors augmented by systematic (model) errors. To enhance robustness, the variance of the residuals was assumed to be unknown and regarded as a mere nuisance parameter. The Gaussian likelihood function was integrated by applying Jeffreys\rq\ prior \citep{Jeff61}. In our case, the improper (not normalisable)
prior varies with the reciprocal of the variance. We  took into account that the binned data points have different weights. This error-integrated likelihood was then integrated analytically over all possible magnitude offsets. 

The resulting likelihood is a function of the data alone, given a set of parameter values. It is a comforting feature of this type of data analysis that the error variance and any magnitude offset, although they remain undetermined, can be ignored, which reduces the number of free parameters and enhances robustness. 

To reduce the number of free parameters as much as possible, we took the star's inclination  to be $i = 36^\circ$. This value results from the measured projected rotational velocity, $v\cdot\sin(i)$, the assumed value of the stellar radius \citep{wire2}, and the rotational period of 4.4790 days \citep{kurtz4}. As the limb-darkening coefficient is unknown, it was regarded as a nuisance parameter for the present study. For the high-quality WIRE data, however, in a follow-up study it might also be possible  to  extract  the marginal distribution of the coefficient for linear limb darkening.

The outcome of a Bayesian parameter estimation is the posterior probability distribution in an high-dimensional parameter space. Direct numerical brute-force integration often requires prohibitively large computing resources, so we used the Markov chain Monte Carlo method (MCMC; cf. \cite{press}). The likelihood mountain in the six-dimensional space is explored by up to 64 Markov chains. In a relaxed state, each parameter's marginal distribution is revealed by counting how often an element in parameter space is frequented by the chains. 

A marginal distribution describes a parameter's probability distribution irrespective of the values of all the other parameters, i.e. disregarding correlations between parameters. These distributions are reflected in Tables\,\ref{tab01} and \ref{tab02} by presenting the corresponding credibility intervals. It is an advantage of the Bayesian approach that parameter averages, such as expectation or median as well as credibility intervals, are deduced from the measured data alone.

\subsection{Surface spot models}    \label{ss:spot}             

We restrict our discussion to two-spot models since the information content of the presently available \b\ data does not enable us to consider more than two spots. The \w\ data do contain more information, which we plan to explore in a future study. The main picture, however, very probably will remain the same.

In our model, two types of circular spots are allowed: dark spots with a brightness of 60\% of the brightness of the undisturbed photosphere and bright spots with a brightness of 125\%. The latter values are similar to what has been reported for the spots of HD\,50773 by \citet{luefta}, based, as is already mentioned, on an {independent} 
analysis of space-based photometry and ground-based Doppler imaging.

The reason for a fixed common spot brightness is the unavoidable degeneracy between spot area and spot brightness, which prevents the MCMC method to converge to a relaxed solution of the posterior probability density distribution. The data simply do not constrain  both parameters independently. Hence, we had no choice but to prescribe the spot brightness parameter, but these are based  on examples in the literature.  
As a test, we repeated the modelling with a $\kappa$ arbitrarily set to 1.5 and expected the inferred spot size to be reduced by about 50\%. Indeed, we found a factor of 1.84, which corresponds to a spot size reduced to 0.54 of the original size. Interestingly, this spot size reduction with increased kappa only holds for the spot that is fully visible. The large spot at the south pole, which basically is visible most of the time, but never entirely, is only reduced by 4\% and its centre moves 6 degrees towards the pole.
It is hardly necessary to stress that it would make even less sense (from the data analysing point of view) to consider spots with individual surface brightnesses. The model spot parameters must be therefore considered with reservation and taken rather as an order-of-magnitude estimate.

The following parameters were estimated for each circular spot: the central longitude ($\lambda$), latitude ($\beta$), and angular radius ($\gamma$). The longitude increases in the direction of stellar rotation, and the zero point is the central meridian facing the observer at the beginning of the time series (HJD = 2456742.188).

Table\,\ref{tab01} and \ref{tab02} summarise what we can say about the locations and sizes of spots on the surface of \ac  based on a Bayesian analysis of the red and blue \b\ light curves and the \w\ light curve. The importance parameter, justifying the attribute ``first" and ``second" to a spot, indicates the impact on the light curve. Listed are {expectation} values  and 90\%  credibility intervals (not to be confused with $\pm 1\,\sigma$ limits). 

\begin{table}
\caption{
Parameters for circular spots for the \b -red, \w, and \b -blue solutions, assuming two {dark} spots. We list expectation values and 90\%  credibility limits for the central longitude ($\lambda$), latitude ($\beta$), and radius ($\gamma$). {Radii and their uncertainties are modelled for a spot intensity of 60\% of the unperturbed photosphere.}
}
\centering
\begin{tabular}{c|rl|rl|rl}
\hline 
Parameter  &\multicolumn{2}{|c|}{\b -red} &\multicolumn{2}{|c|}{WIRE}& \multicolumn{2}{|c}{\b -blue}   \\ [4pt]
\hline  
\multicolumn{7}{c}{first spot}   \\[4pt]
$\lambda$  &$  74^\circ\hspace{-3pt}.3$ &\hspace{-10pt}$^{+ 2.4}_{- 2.4}$ &${ 89^\circ\hspace{-3pt}.4 }$ &\hspace{-10pt}$^{+0.8}_{-0.7}$ &\hspace{10pt}$ 209^\circ$ &\hspace{-10pt}$^{+24}_{-29}$  \\[4pt]%
$\beta$       &$ -72^\circ\hspace{-3pt}.7$ &\hspace{-10pt}$^{+ 5.5}_{- 5.2}$ &${ -2^\circ\hspace{-3pt}.8 }$ &\hspace{-10pt}$^{+3.4}_{-3.7}$ &                      $  36^\circ$   &\hspace{-10pt}$^{+41}_{-33}$  \\[4pt]%
$\gamma$  &$  58^\circ\hspace{-3pt}.2$ &\hspace{-10pt}$^{+ 6.2}_{- 6.3}$ &${  8^\circ\hspace{-3pt}.0 }$ &\hspace{-10pt}$^{+0.4}_{-0.3}$  &                      $   6^\circ$   &\hspace{-10pt}$^{+ 2}_{- 2}$     \\[4pt]%
\multicolumn{7}{c}{second spot}   \\[4pt]%
$\lambda$  &$ 248^\circ\hspace{-3pt}.5$ &\hspace{-10pt}$^{+ 6.2}_{- 6.2}$ &${263^\circ\hspace{-3pt}.4 }$ &\hspace{-10pt}$^{+1.3}_{-1.3}$  &$ 118^\circ$ &\hspace{-10pt}$^{+12}_{-14}$ \\[4pt]%
$\beta$       &$   6^\circ\hspace{-3pt}.0$  &\hspace{-10pt}$^{+44.8}_{-30.8}$&${-18^\circ\hspace{-3pt}.7 }$  &\hspace{-10pt}$^{+2.9}_{-3.1}$  &$ -24^\circ$ &\hspace{-10pt}$^{+25}_{-35}$ \\[4pt]%
$\gamma$  &$   6^\circ\hspace{-3pt}.0$  &\hspace{-10pt}$^{+ 1.8}_{- 2.2}$  &${  8^\circ\hspace{-3pt}.3 }$   &\hspace{-10pt}$^{+0.7}_{-0.7}$  &$  16^\circ$ &\hspace{-10pt}$^{+15}_{-11}$ \\[4pt]%
\hline
residuals     &\multicolumn{2}{|c|}{$\pm$ 2.1}   &\multicolumn{2}{|c|}{$\pm$ 0.14}  &\multicolumn{2}{|c}{$\pm$ 4.8} \\ 
\hline
\end{tabular}
\label{tab01}
\end{table}

\begin{table}
\caption{The same as Table\,\ref{tab01}, but assuming  {bright} spots. No two-spot solution was obtained for the \b -blue data. {Radii and their uncertainties are modelled for a spot intensity of 125\% of the unperturbed photosphere.}
} 
\centering
\begin{tabular}{c|rl|rl|rl}
\hline 
Parameter  &\multicolumn{2}{|c|}{\b -red} &\multicolumn{2}{|c|}{WIRE}& \multicolumn{2}{|c}{\b -blue}   \\ [4pt]
\hline  
\multicolumn{7}{c}{first spot}   \\[4pt]
$\lambda$  &$ 326^\circ\hspace{-3pt}.5$ &\hspace{-10pt}$^{+ 7.8}_{- 7.4}$   &${356^\circ\hspace{-3pt}.9 }$ &\hspace{-10pt}$^{+0.9}_{-0.8}$ &\hspace{10pt}$ 356^\circ$ &\hspace{-10pt}$^{+ 8}_{- 8}$  \\[4pt]%
$\beta$       &$  31^\circ\hspace{-3pt}.7$  &\hspace{-10pt}$^{+25.1}_{-21.9}$&${ 25^\circ\hspace{-3pt}.9 }$  &\hspace{-10pt}$^{+4.3}_{-4.4}$  &                      $  34^\circ$ &\hspace{-10pt}$^{+27}_{-23}$\\[4pt]%
$\gamma$  &$   9^\circ\hspace{-3pt}.6$   &\hspace{-10pt}$^{+ 2.0}_{- 2.0}$  &${ 13^\circ\hspace{-3pt}.7 }$  &\hspace{-10pt}$^{+1.4}_{-1.4}$  &                      $   8^\circ$  &\hspace{-10pt}$^{+ 1}_{- 1}$\\[4pt]%
\multicolumn{7}{c}{second spot}   \\[4pt]%
$\lambda$  &$ 172^\circ\hspace{-3pt}.2$ &\hspace{-10pt}$^{+ 7.3}_{- 7.6}$    &${186^\circ\hspace{-3pt}.1 }$ &\hspace{-10pt}$^{+2.0}_{-2.0}$   && \\[4pt]%
$\beta$       &$ -73^\circ\hspace{-3pt}.8$  &\hspace{-10pt}$^{+ 5.2}_{- 5.3}$&${-82^\circ\hspace{-3pt}.8 }$ &\hspace{-10pt}$^{+1.4}_{-1.5}$       &\multicolumn{2}{c}{no solution} \\  [4pt]%
$\gamma$  &$  64^\circ\hspace{-3pt}.2$  &\hspace{-10pt}$^{+ 7.3}_{- 7.2}$&${ 85^\circ\hspace{-3pt}.2 }$ &\hspace{-10pt}$^{+4.4}_{-3.6}$        && \\[4pt]%
\hline
residuals     &\multicolumn{2}{|c|}{$\pm$ 2.1}   &\multicolumn{2}{|c|}{$\pm$ 0.13}  &\multicolumn{2}{|c}{$\pm$ 4.9} \\ 
\hline
\end{tabular}
\label{tab02}
\end{table}

A comparison of the tables shows that dark and bright spots individually fit the \bc\ and \w\ data equally well, i.e. we find similar credibility limits. The model light curves (full lines in Fig.\,\ref{comcol}) look very much the same and give the same residuals to the observations, regardless of whether the spots are dark or bright. However, important differences come to light in a detailed examination.

The surface mapping for two dark circular spots with the red \b\ and \w\ data requires a separation of 180$^\circ$ in longitude with spots on opposite sides of the star. Two dark spots, which fit the blue \b\ data, however, are separated by nearly 90$^\circ$ in longitude and do not agree with the red map. This suggests that bright spots are a more realistic fit to the entire data set than dark spots. 

\begin{figure}[t]
\begin{center}
\resizebox{\hsize}{!}{\includegraphics{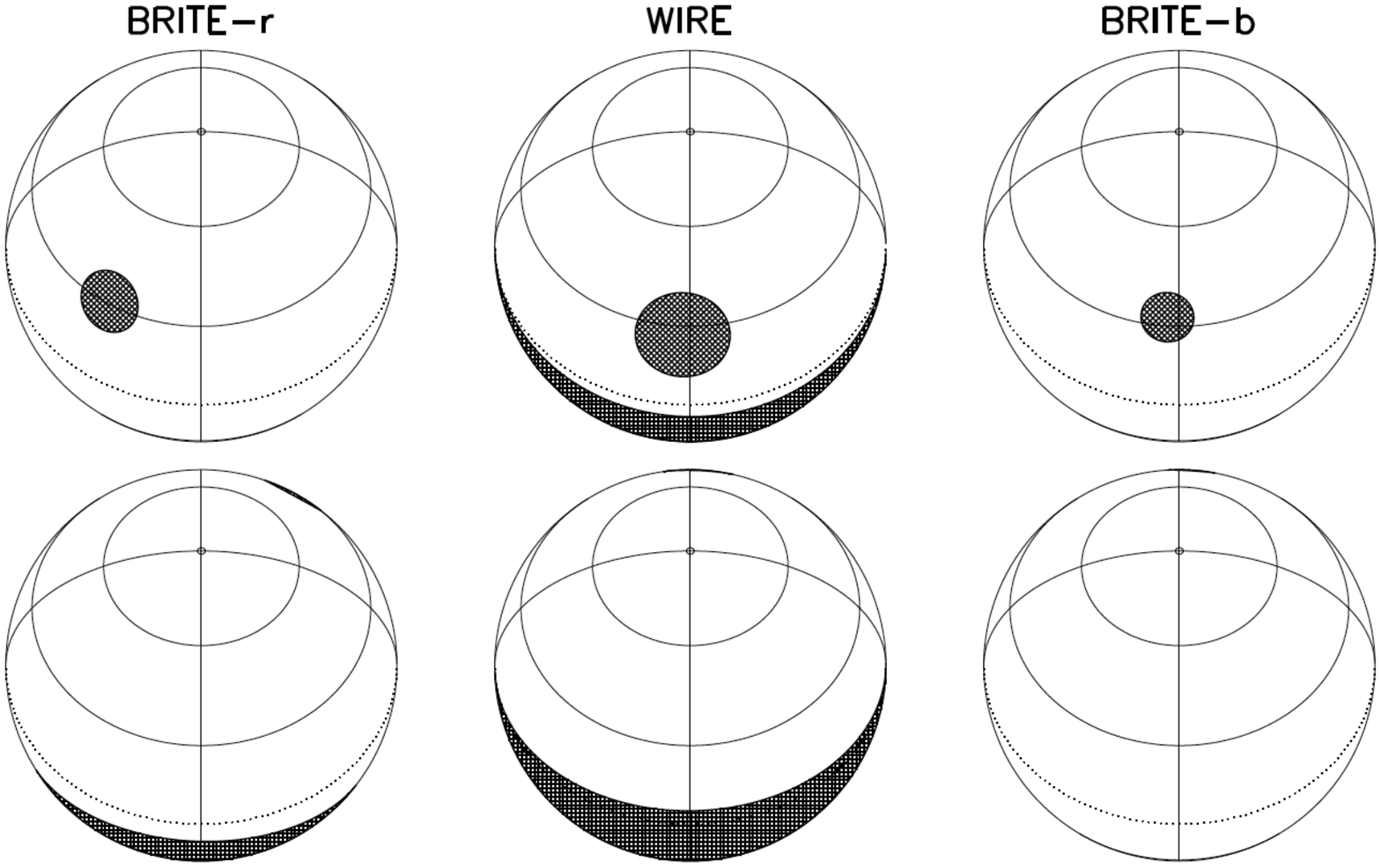}}
\caption{Maps of bright spots based on \b \ blue, \b \ red, and \w\  according to the parameters given in Table\,\ref{tab02}. Two rotation phases shown (top and bottom) differ by 180\deg . The Bayesian technique cannot identify   a spot at the stellar south pole  with sufficient probability for the \b -blue data, probably because of a lower photometric accuracy.} 
\label{bpi}
\end{center}
\end{figure}

We only find    a spot model that agrees with \w\  and both \b\ data sets in the case of bright spots. However, only one bright spot could be detected in the \b\ blue data, which is not surprising, considering the differences in the red and blue \b\ light curves (see Fig.\,\ref{comcol}). 
A schematic of what we believe is the most realistic surface model for \ac\ (see Table\,\ref{tab02}) is provided in Fig.\,\ref{bpi}. The spot close to the stellar south pole is credibly present in the \w\ and \b-red data, but not in the \b-blue data, which have the lowest precision. 

An inhomogeneous surface distribution of elements, like Si, Cr, and Fe, are known to cause variations in the light curves of many Ap stars detectable as bright spots in visual photometry \citep[e.g.][]{luefta,lueftb,krt0,krt1,krt2,shul}. This is a strong argument to favour bright spot solutions over dark spot solutions, if no other information is available. The size of abundance spots  is independent of wavelength, which is what we find for the bright spot model within twice the credibility limits.
%

Bayesian techniques provide an infinite number of feasible two-spot solutions, but these differ in credibility. A main problem in our case is the skewness of the marginal distribution for the parameters, in particular, for the \b\ data, as is evident from the asymmetric 90\% credibility intervals, which is also listed in both tables. The focus of our investigation is on the expectation values of the parameters\rq\ marginal distributions. Such a spot model is, in a certain sense, the most representative, but because of the posterior\rq s skewness, this  model may differ from the globally most probable model (the \lq
best-fit model\rq ).

\subsection{Significance of BSI models} 
In  the previous Section we tried to model the observed light curves obtained in different colours with circular areas at the stellar surface (spots), which manifest themselves to the observer as periodically varying {disk averages} obtained at various {rotation phases}. Stellar surface structures can be very complicated because they depend on temperature, gravity, dynamics in the atmosphere, element abundances, and magnetic field,  all  of these as a functions of optical depth. In addition, insufficient knowledge of atomic and molecular transition properties and energy level splittings in the presence of a magnetic field can have serious effects on how such surface areas are interpreted by observers.

The transformation of a one-dimensional function, the light curve, into a two-dimensional surface map obviously requires simplifying assumptions. An adjustable model constrained by disk-averaged flux measurements, obtained with three broadband filters at a discrete set of rotation phases, cannot represent the complex physics and chemistry of a stellar atmosphere, but can indicate locations of inhomogeneities. Unfortunately, in the present investigation we only have  a limited amount of information available for \ac\ to map the complex surface. Photometric surface models cannot recover information about  properties such as abundances and temperature, but they can indicate areas  with inhomogeneities at the stellar surface. In addition from consistency arguments, they can indicate a preference for flux excess or depression with no reference to an underlying physics.  Therefore, a realistic starting point  is a model, which is as simple as possible and is tested against the observations, which in our case is light curves. 

The Bayesian approach offers two attractive features: (a) the {quantitative} comparison of models differing in the number of spots and (b) the derivation of {credibility} intervals for the model parameters from the posterior probability distribution. Our two-spot model is sufficient to reproduce the photometry with the best  probability distribution possible so far. Adding more features may suggest a better visual representation of the light curve, but considering the noise level of the data, increased complexity reduces the over-all credibility of the model. Indeed, we have investigated more complex models (e.g. three spots, different limb darkening coefficients, and spots of different brightness), but had to accept that all these additional model features did not increase the credibility. Moreover, with too many free parameters involved, the method does not work because the Markov chains do not converge to a well-constrained posterior probability density distribution. The only remedy are additional observations that access new parameter space, such as high spectral and time-resolved spectroscopy and/or spectropolarimetry. 
{The limited amount of presently available data, their quality and physical information content simply do not allow for a more sophisticated spot modelling, for example for disentangling $\kappa$ and size of spots.}

Nevertheless, our present model serves as a reasonable starting point for more sophisticated modelling of the surface structure, for example based on future analysis of spectroscopic and/or spectropolarimetric data. This expectation is supported by the results of  \citet{luefta} for HD\,50773, where essentially the same surface pattern was determined independently from Bayesian spot modelling of space-based photometry (CoRoT) and from Doppler imaging modelling of spectropolarimetric data. Even more significantly, \citet{alen} report surface abundance maps of Cr\,II in their preliminary Doppler imaging analysis,  which support our model (= course map of inhomogeneities) derived from photometry.

\section{Pulsation frequencies, amplitudes, and phases}    

The pulsation spectrum of \ac\ shows a dominant mode with a period of $\sim6.5$\,min ($f_{1}$), which is rotationally split into a triplet by rotation \citep{kurtz4}. Four additional low-amplitude frequencies ($f_{2}$ to $f_{5}$), attributed to roAp oscillations, were reported by \citet{kurtz4}. Using \w\ data, \citet{wire1} confirmed two of these low-amplitude modes ($f_{4}$ and $f_{5}$), while $f_{2}$ and $f_{3}$ remained undetected. 

\citet{wire1} also reported detection of two new frequencies ($f_{6}$ and $f_{7}$) that are symmetrical to $f_{1}$, which had not been seen in previous ground-based campaigns nor in the \w\ 2000 and 2005 data. 

\subsection{Frequencies}

\begin{figure*}
\begin{center}
\resizebox{\hsize}{!}{\includegraphics{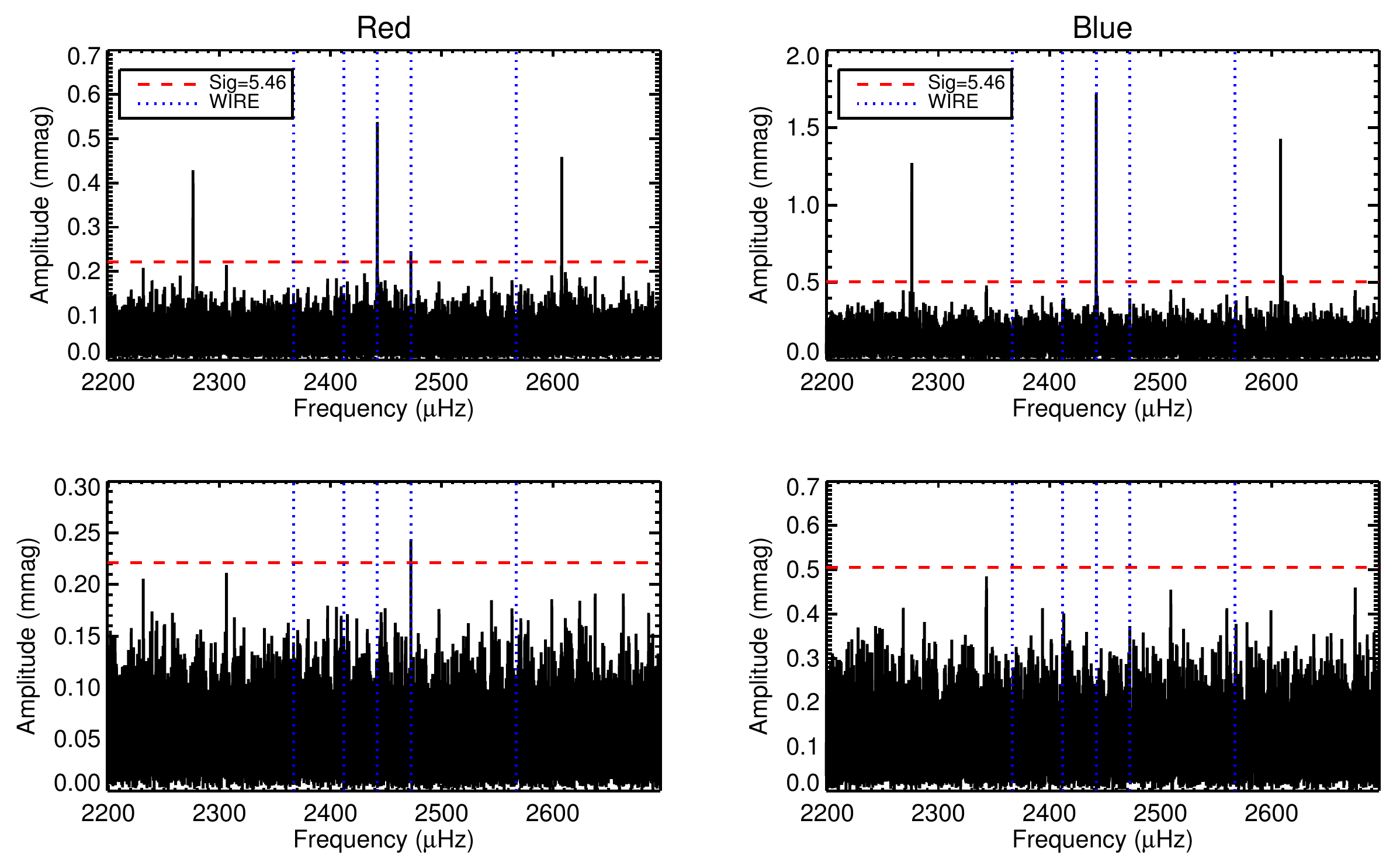}}  
\caption{Amplitude spectrum of the red (left panels) and blue (right panels) \b\ light curves after removing rotational variability and instrumental effects. The top panels show amplitude spectra of the  data, while the bottom panels show amplitude spectra after pre-whitening of the dominant pulsation mode near $\sim 2442\,\muHz$. The sidelobes seen in the top panel correspond to the orbital frequency of the satellites. Vertical dotted lines in each panel are published \w\ frequencies by \citet{wire1}, while the red horizontal lines show a spectral significance corresponding to a false-alarm probability of $10^{-5.46}$ (or S/N$\sim 4$).}
\label{fig:puls1}
\end{center}
\end{figure*}

We first removed the contribution of rotational variability and instrumental effects (1\cd , 2\cd , orbital frequency and its harmonics) from the light curves with a multi-sine fit provided by Period04 \citep{lenz}. Next, we calculated amplitude spectra and estimated the statistical significance of frequencies via SigSpec \citep{reegen07,reegen}, which has previously been extensively applied to space-based photometry from MOST, including roAp stars \citep{gruberbauer08,huber08}. Figure \ref{fig:puls1} shows the amplitude spectra of the red and blue \b\ data centred on the frequency range where oscillations had already been observed in \ac . The dominant mode near $\sim6.5$\,min ($\sim 2442\,\muHz$) is clearly detected in both data sets. As expected, the amplitude is significantly higher in the blue than in the red filter.

\begin{figure}
\begin{center}
\resizebox{\hsize}{!}{\includegraphics{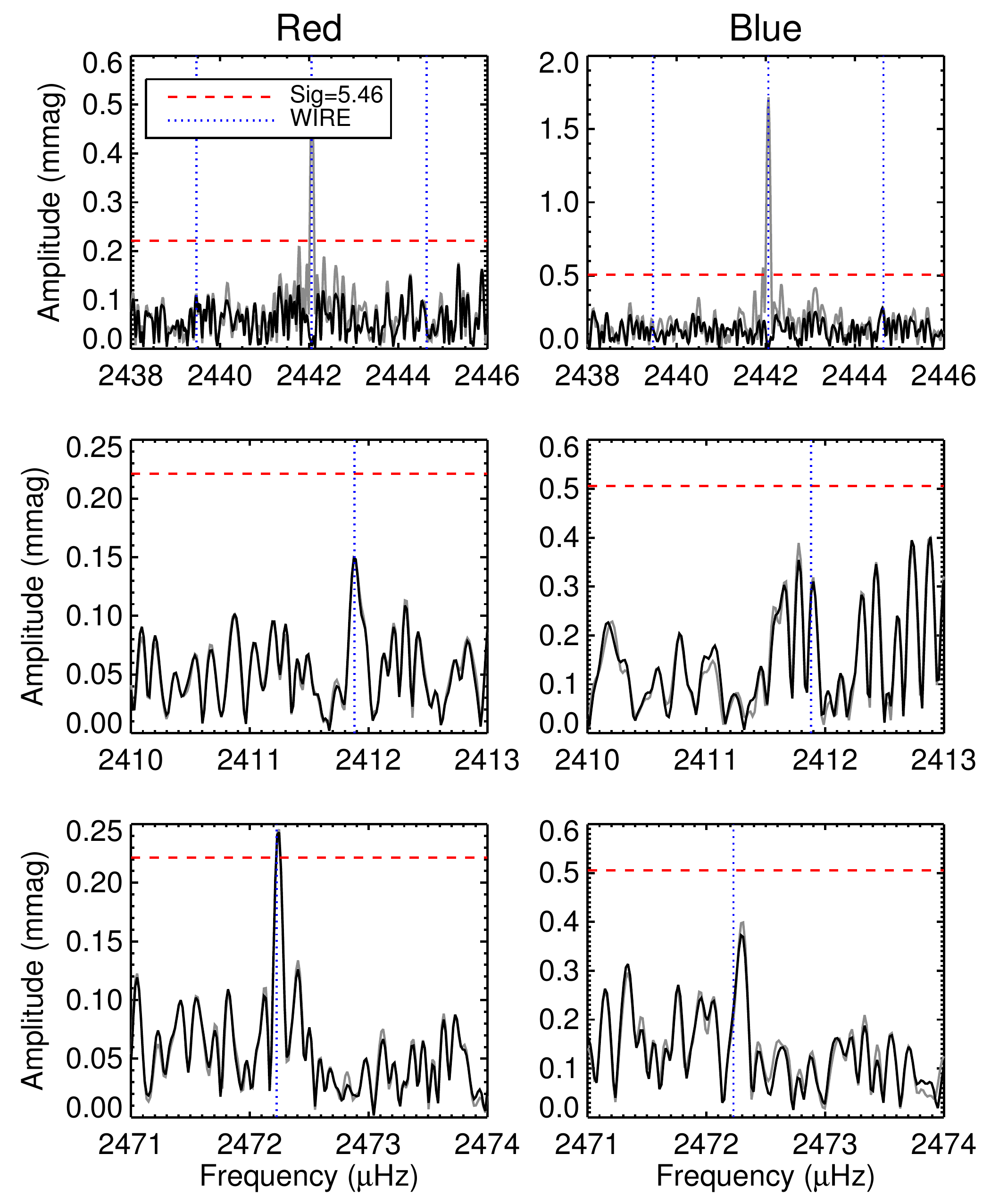}}    
\caption{Same as Figure \ref{fig:puls1} but for a close up of the rotational multiplet around $f_{1}$ (top panels) as well as $f_{6}$ (middle panels) and $f_{7}$ (bottom panels). Grey and black lines show the spectrum before and after pre-whitening of $f_{1}$, respectively.}
\label{fig:puls2}
\end{center}
\end{figure}

The top panels of Fig.\,\ref{fig:puls2} show a close-up of $f_{1}$ demonstrating that the rotationally split side lobes remain undetected in the \b\ data. This is not due to the removal of the rotational variability from the light curve, since the multiplet is caused by rotational amplitude modulation of $f_{1}$ as a consequence of the oblique pulsator model \citep{kurtz86}. The apparent contradiction of an $f_1$ amplitude changing with rotation period and the lack of rotation side lobes seen in the frequency spectrum of our data is due to very small amplitudes. \citet{kurtz4} estimated these side-lobe amplitudes to be about one-tenths of the main amplitude (=\,0.55\,mmag for $f_{1}$ and our red \b s ) and give a justification for this value within the framework of the oblique pulsator model. 

The middle and bottom panels of Fig.\,\ref{fig:puls2} show the amplitude spectra after pre-whitening the data by the dominant mode $f_{1}$ and we detect $f_{7}$ in the \b-red data, while in both data sets $f_{6}$ remains below the detection threshold (S/N ratio of $\sim 4$). This ratio corresponds to a significance limit of $10^{-5.46}$, or to a 99.9\% detection probability of a true signal \citep{kusch}. The amplitude ratios are A($f_{1}$)/A($f_{7}$) = 3.468($\pm 0.007$) for \w\ and 2.51($\pm 0.10$) for \b\  red. \citet{wire1} speculate that $f_{6}$ and $f_{7}$ first appeared between 2005 and 2006 and interpreted them as consecutive overtones of a new pulsation mode. Unfortunately, the noise level of the blue \b\ data does not enable a detection of $f_{7}$, which would be helpful in discussing the pulsation mode. It has to be proven by follow-up observations that indeed a mode that is {different} in parity than $f_{1}$ has been excited. A new mode of the {same} parity would have produced a frequency separation that is twice as large (see discussion in \citet{wire1}).

Using the frequencies identified with SigSpec, we applied Period04 \citep{lenz} to calculate least-squares fits for frequency, amplitude, and phase to each light curve. For $f_{1}$ we fixed the frequency to the best-fit values from the blue data, which has higher significance than the red data. Uncertainties on all values were calculated with Monte Carlo simulations, which are implemented in Period04. The final values, uncertainties, and significances  are listed in Table \ref{tab:puls}.

\subsection{Amplitudes and phases}

The pulsation amplitude ratio for $f_{1}$ measured from \b\ observations is $A_{B}/A_{R}=3.15\pm0.33$, compared to $A_{B}/A_{WIRE}=2.32\pm0.11$ by \citet{wire1} and $A_{B}/A_{V}=2.28\pm0.26$ by \citet{kurtz3}. Formally, all values agree within $2\,\sigma$, and the \b\ amplitude ratio seems to confirm the conclusion by \citet{wire1} that the \w\ bandpass is likely close to the $V$ band. 

The mentioned pulsation amplitude ratio is consistent with the relation among photometric amplitudes in different photometric bands known from previous ground-based observations, i.e. amplitude decreases with increasing wavelength \citep[e.g.][]{weiss1,kume}. 

Considering the amplitude ratios for $f_{1}$ between \b\ and \w , the expected amplitudes for $f_{4}$ to $f_{6}$ found by \w\ would be 0.04 to 0.12\,mmag in the red filter and  0.12 to 0.40\,mmag in the blue filter. Given a typical noise of $\sim 0.06$ \,mmag for red \b\ and $\sim 0.13 $\,mmag for blue \b\ data, we conclude that our non-detections are consistent with the \w\ results by \citet{wire1}. However,  the amplitude ratio for $f_{1}$ may not necessarily apply for $f_{4}$ and $f_{5}$, which must have a different spherical degree than $f_{1}$, $f_{6}$ and $f_{7}$, if the large frequency separation claimed by Bruntt et al. (op.cit.) is correct.


\citet{kurtz3} and \citet{wire1} noticed a slight phase shift of $\phi_{V}-\phi_{B}=-7.4\pm5.1^{\circ}$ and $\phi_{WIRE}-\phi_{B}=5.8\pm1.3^{\circ}$. Using the pulsation phases calculated from the unbinned blue and red \b\ light curves (calculated at a fixed frequency $f_{1}$),  the \b\ observations yield a phase shift of $\phi_{R}-\phi_{B}=10.6\pm5.9^{\circ}$, which is consistent with previous results.

\section{Discussion}        \label{s.disc}

{\sc{-- Rotation:}}    
Our \b\ red data (similar to a Johnson R bandpass) confirm the rotation period of 4.4790\, days and the light curve characteristics of \ac\ determined by \w\ (similar to Johnson V), showing two maxima at about $\phi$ = 0.2 and 0.8 in our Fig.\,\ref{comcol}. This agreement is contrasted by our \b\ blue data (similar to Johnson B), which show only one maximum at about $\phi$ = 0.7. 

This profound difference in the shape of light curves obtained in different filters has not been seen before and would normally cast doubts on the reliability of our blue data. However, strong support is provided by: \\ 
\hspace*{2pt}i) RV measurements which are consistent with a rotation phase plot having only one maximum  (Mkrtichian \& Hatzes, private communication);  \\ 
ii) magnetic field measurements, which also indicate a single-wave variation of the magnetic field (Kochukhov, private communication).  

It is a known property of CP (Ap) star rotation light curves that amplitudes depend on the photometric filter used (e.g. HD\,30849: \citet{hens}). Even maxima and minima can switch (e.g. HD\,53116, op.cit.) when switching colours (more examples in \citet{schoen}), but the disappearance of only one extremum and not also the other is unique. Other references to changes of the shape of light curves with wavelength are, e.g. \citet{musi,kuma,miku1,miku2}. 

\begin{table}
\begin{small}
\caption{$\alpha$\,Cir pulsation frequencies extracted from \b\ data}
\begin{center}
\begin{tabular}{l c c c c}
\hline
Data &  Frequency & Amplitude & Phase & Sig \\
         &      (\muHz) & (mmag) & (0-1) &  \\
\hline
\multicolumn{5}{c}{Frequency $f_{1}$} \\
red \b\ & [\ \  2442.0574\ \ ]  & 0.545(47)     & 0.191(14)     &       33.2    \\
blue \b\        & 2442.0574(31) & 1.72(10)      & 0.165(09)     &       63.1    \\
\hspace{10pt}{\em WIRE}   & {\em 2442.0566(2)}   & {\em 0.652(5)}     \\
\hline
\multicolumn{5}{c}{Frequency $f_{7}$} \\
red \b\ & {2472.2401(88)} & {0.249(43)} & {0.057(32)}   &       {6.8}   \\
blue \b\        & -- & -- & -- & --     \\
\hspace{10pt}{\em WIRE}   & {\em 2472.2272(5)}   & {\em 0.188(5)}     \\
\hline                                                                                               
\end{tabular} 
\end{center}
\flushleft Notes: 
For the red \b\ data ($\approx$Johnson R), we fixed $f_{1}$ to the best-fit value [in parentheses] from the blue \b\ data ($\approx$Johnson B) since the pulsation amplitude is larger in blue and has higher significance, despite the larger noise level in blue.
The last column lists the spectral significance derived with SigSpec. Uncertainties are calculated via Monte Carlo simulations  implemented in Period04. 
The reference for \w\ ($\approx$Johnson V) is Table 3 of \citet{wire1}.
\label{tab:puls}
\end{small}
\end{table}

Abundance anomalies on the surfaces of magnetic CP stars, which imprint their signatures on the light curves and spectral line profiles, produce changing redistributions of stellar flux (primarily from short to longer wavelengths). 
This effect, which is still not well understood, might lead to  considerably different shapes of  \ac\ light curves in the blue and red bands observed by \bc . \\ [5pt]
{\sc{ -- Spots:}}    
Current knowledge about flux redistribution results in dark spots in short-wavelength regimes, but bright spots in the optical, if the spots are due to concentrations of elements like Si, Cr, or Fe \citep{krt1}. 
\citet{lueftb}, Shulyak et al. (2010), and \citet{krt2} find that variations of HD\,37776 and $\theta$\,Aurigae in the $u, v, b, \rm{and}\ y$ Str\"omgren bands are caused by spots dominated by Si, Cr, and/or Fe, and those of HD\,37776 are dominated by He. The elements Fe and Cr are primarily responsible for the variability of $\epsilon$\,UMa \citep{shul}.

Considering the evidence provided by our red and blue \b\ light curves, it is not surprising that the derived spot models are unusual. We speculate that the second spot has  a very different chemical composition than the first spot, which results  in an equilibrium of redistributed and absorbed flux for the blue filter. 

\citet{alen} derived very preliminary spectroscopic surface maps for \ac\ with spots dominated by Si and Cr, and detected abundance variations of up to 2.4\,\,dex, which support our model solution assuming bright spots.\\ [5pt]
{\sc{ -- Pulsation:}}   
The dominant { frequency} near 6.5\,min (2442\,\muHz) is clearly detected in both \b\ data sets. 
With a SigSpec significance of 6.8 (S/N > 4), we find also $f_7$ in the frequency spectrum of the red \b\ data (Fig.\,\ref{fig:puls2}, bottom). However,  $f_{7}$ does not appear in our blue data, even if we lower the threshold to a false-alarm probability of 0.1\% (S/N$\sim$3).  According to the amplitude
scaling with wavelengths, however,  $f_{7}$
should appear in our blue data, if it is  related to the pulsation mode of $f_1$. We emphasise that $f_{7}$ has only been detected since 2006.

The pulsation {amplitude} of $f_1$ decreases from the blue to red wavelengths (see our Table\,\ref{tab:puls}), a trend that is well known for roAp stars. Since the amplitude decline rate (from Johnson B to R filters) is much steeper for pulsating late A-type stars than that of main-sequence B-type variables, \citet{med} explained the case of \ac\ with an outwards decreasing temperature amplitude within a narrow range ($\sim\,200$\,km) of the atmosphere.

The observed {phase shift} between the blue and red \b\ data of $10^{\circ}\pm6^{\circ}$ is consistent with previous determinations.
This small phase difference between red and blue light curves is in sharp contrast to considerable phase variations in the radial velocity oscillations of roAp stars, obtained from various spectral lines and at various depths of the H$\alpha$ line \citep{bal1,bal2, ryab}.  Such spectroscopic studies have shown that the pulsation properties vary significantly with  atmospheric depth due to chemical stratification, leading to phase shifts between different elements.

\begin{figure}[]
\begin{center}
\includegraphics[width=0.35\textwidth]{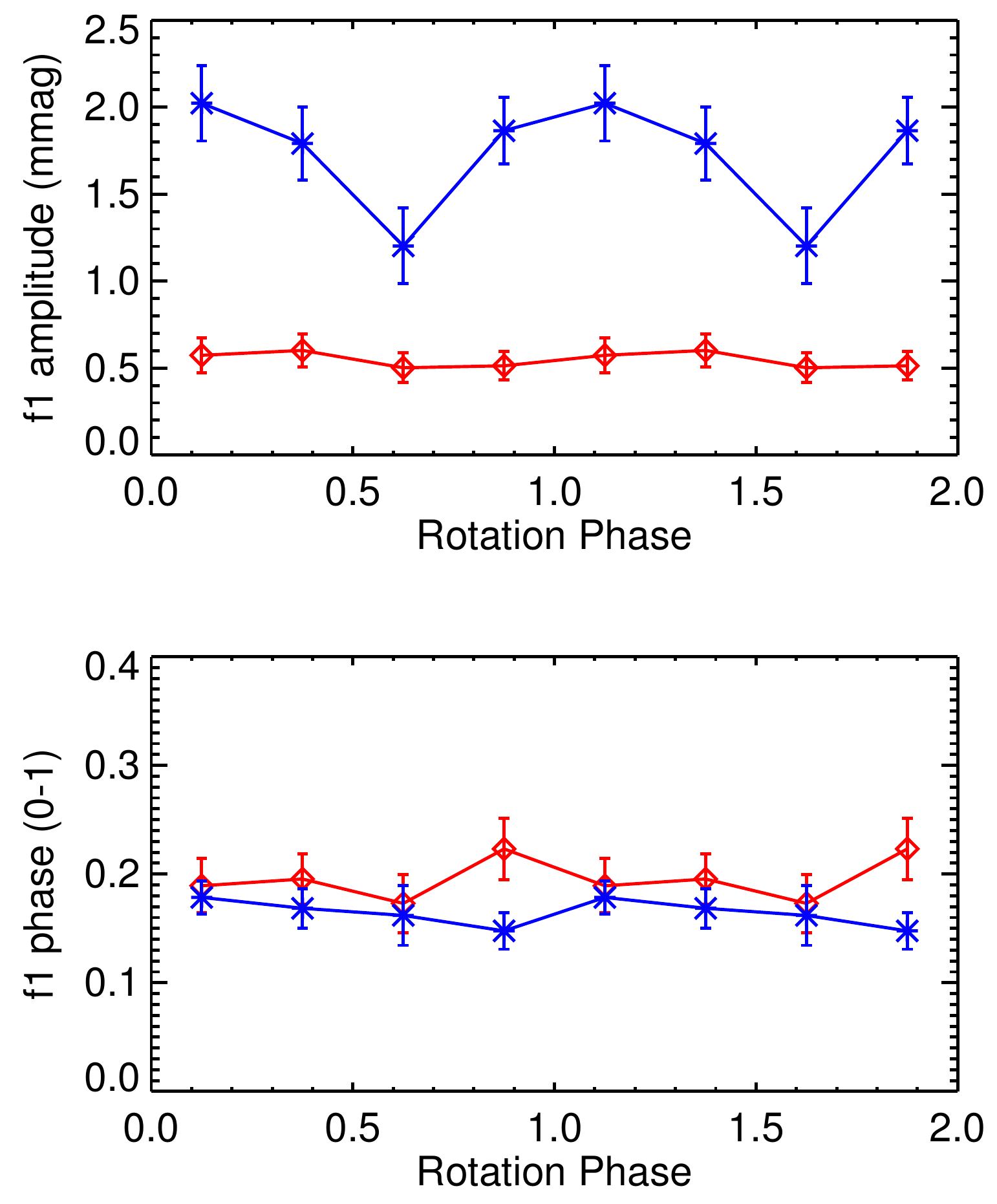}
\caption{Dependence of amplitude and phase of frequency $f_1$ on the rotation phase of \ac , observed by blue (crosses) and red (dots) \b s.}
\label{fig:amphas}
\end{center}
\end{figure}

Pulsation energy, generated below the photosphere, flows outwards into the atmosphere with changing phases. This phase variation is faster in the outer parts of the atmosphere because some of the energy leaks at the outer boundary, where the non-adiabatic damping effects are stronger. On the other hand, the phase variation is slow at the base of the atmosphere because non-adiabatic effects are weak. Hence, RV variations deduced from rare earth elements and H$\alpha$ bisectors represent pulsation in the outer part of the atmosphere. There, non-adiabatic damping  is strong, while the photometric variations primarily represent  temperature effects due to pulsation near the bottom of the photosphere (optical depth $\tau\approx 1$), where non-adiabatic effects are expected to be weak. 

All the evidence points to high-order p-mode pulsation in \ac . We refer to Fig.\,6 of \citet{wire1} where the authors discuss amplitude changes for various axisymmetric pulsation modes depending on the rotation phase of \ac . Our Fig.\,\,\ref{fig:amphas} compares amplitude and phase changes of $f_1$ observed in the red and blue by \b\ nanosats, averaged over 25\% of a full rotation cycle, respectively; our figure
also addresses Fig.\,10 of \citet{kurtz4}.  Our conclusion is that within the quoted errors, quadrupolar modes can be excluded for $f_1$, and {$\ell\  = 1, m = 0$} being supported.

Investigations of long-term changes of rotation and pulsation properties are obvious next research goals. Right now the time base of high quality data is too short, but observers are encouraged to monitor \ac , as  \bc\ will as well.
\\ [5pt]
{\sc{ -- Final note:}}   
High-resolution spectroscopy with high S/N, covering a full rotation cycle, is the obvious strategy to test our two bright-spot model and its dependence on wavelengths. Spectroscopy will  also be critical to investigate chemical and physical spot characteristics as a function of optical depths and spectroscopy with high time resolution will be required to determine the stellar structure anomalies. 

In other words, having information both on the temperature and RV variations in the atmosphere with good estimates of fundamental parameters (e.g. \citet{wire2}), \ac\ is one of the best roAp stars to constrain  the model of pulsation in the presence of a strong magnetic field and to look deep into a star.

\begin{acknowledgements}
RK and WW were supported by the Austrian Science Funds (FWF) and by the Austrian Research Promotion Agency (FFG-ASAP), which also supported OK. AFJM is grateful for financial assistance from NSERC (Canada) and FRQNT (Quebec). JMM, SMR, and GAW are grateful for support from NSERC (Canada). The Polish BRITE operations are funded by the PMN grant 2011/01/M/ST9/05914.
APi acknowledges the support from the NCN grant No. 2011/03/B/ST9/02667 and APo was supported by the Polish National Science Center, grant no. 2013/11/N/ST6/03051: Novel Methods of Impulsive Noise Reduction in Astronomical Images. The MCMC computations have been performed by HEF at the AIP. GH and APa received support by the Polish NCN grant 2011/01/B/ST9/05448. DH acknowledges support by the Australian Research Council's Discovery Projects funding scheme (project number DE140101364) and support by the National Aeronautics and Space Administration under Grant NNX14AB92G issued through the Kepler Participating Scientist Program. TL acknowledges funding of the Austrian FFG within ASAP11 and support by the FWF NFN projects S11601-N16 and S116 604-N16. K.Z. acknowledges support by the Austrian Fonds zur Fo\"rderung der wissenschaftlichen Forschung (FWF, project V431-NBL). 
Finally, the authors wish to acknowledge the spacecraft operation teams in Austria (P. Romano \& M. Unterberger), Canada (Monica Chaumont, Susan Choi, Daniel Kekez, Karan Sarda, Paul Choi \& Laura Bradbury) and Poland (Grzegorz Marciniszyn \& Grzegorz Wo\'zniak), whose efforts were essential for the collection of the data used in this paper.

\end{acknowledgements}


\end{document}